\documentclass[twocolumn,showpacs,amsmath,amssymb,nofootinbib]{revtex4}

\usepackage{graphicx}
\usepackage{dcolumn}
\usepackage{bm}

\newcommand{\be}{\begin{equation}}
\newcommand{\en}{\end{equation}}
\newcommand{\bea}{\begin{eqnarray}}
\newcommand{\ena}{\end{eqnarray}}

\begin{document}

\preprint{GACG/02/2005}

\title{ Curvaton reheating in tachyonic braneworld inflation }

\author{Cuauhtemoc Campuzano}
 \email{ccampuz@mail.ucv.cl}
\affiliation{ Instituto de F\'{\i}sica, Pontificia Universidad
Cat\'{o}lica de Valpara\'{\i}so, Casilla 4059, Valparaiso, Chile.}
\author{Sergio del Campo}
 \email{sdelcamp@ucv.cl}
\affiliation{ Instituto de F\'{\i}sica, Pontificia Universidad
Cat\'{o}lica de Valpara\'{\i}so, Casilla 4059, Valparaiso, Chile.}
\author{Ram\'on Herrera}
\email{rherrera@unab.cl} \affiliation{Departamento de Ciencias
F\'\i sicas, Universidad Andr\'es Bello, Avenida Rep\'ublica 237,
Santiago, Chile.}

\date{\today}

\begin{abstract}
The curvaton reheating in a tachyonic braneworld inflationary
universe model with an exponential potential is studied. We have
found that the energy density in the kinetic epoch, has a
complicated dependencies of the scale factor. For different
scenarios the temperature of reheating is computed, finding an
upper limit that lies in the range $10^{14}$--$10^{16}$GeV.
\end{abstract}

\maketitle
\section{\label{sec:level1} Introduction}

Inflationary models predict a flat universe and a nearly
scale-invariant spectrum of primordial density perturbations
\cite{prima}. One of the predictions of inflation that needs to be
tested is a stochastic gravitational - wave background with a
nearly scale-invariant spectrum \cite{ruvakov}. In this way,
detection of the cosmic microwave background polarization pattern
induced by inflationary gravitational waves of wavelengths
comparable to the horizon has become a goal of next-generation CMB
experiments \cite{kamion}.

During the inflationary period the universe becomes dominated by
the inflation scalar potential. But at the end of this phase the
universe presents a combination of kinetic and potential energies
associated to this scalar field, which is assumed to occur at very
low temperature. The way of defrosting the universe after the
period of inflation is know as reheating. Theory of reheating for
the new inflationary universe was studied in Refs.\cite{3} .
During this period, most of the matter is created via the decay of
the inflaton scalar field, while the temperature grows at high
enough value, such that the standard Big-Bang universe model is
recovered.

All this process occurs due to collisions and decay in which the
quasiperiodic evolution of the scalar fields lead to a creation of
particles and thus the universe becomes hot. How hot results the
universe depend on the quantity known as the reheating
temperature, which is defined by assuming a conversion of energy
density associated to the scalar field into radiation, while the
decay width of the energy density happens.

In the standard theory of reheating the period of oscillation of
the inflation field is a crucial part. These oscillations occur at
the minimum of the potential. However, there exist some models
where the inflation potential does not have a minimum and thus the
scalar field could not oscillate. These sort of models are known
as non-oscillating (NO) models \cite{Felder1}. The implementation
of reheating in these kind of scenarios have been varied: The
"gravitational particle production" \cite{ford}, which has been
shown to be inefficient, since it may lead to some cosmological
problems \cite{urena,Sami_taq}. As an alternative mechanism of
reheating is the "instant reheating", which characterizes by
introducing an interactions between the inflation field and an
additional scalar field. These model depends strongly on the
parameters introduced via the interaction term. Perhaps, the most
efficient form of implementation reheating in NO models is the
so-called ``curvaton reheating'' \cite{ref1u}, which consists on
introducing a new scalar field, apart of the inflaton the so
called curvaton field. At difference of the instant reheating
case, there is not interaction between these two scalar fields
\cite{fengli}. The main characteristic of this scenario is that
the matter described by the scalar curvaton is such that the
energy density is not diluted during inflation and some or all of
the present matter in the universe may have survive inflation via
the curvaton scalar field.

On the other hand, the idea of considering extra dimension has
received great attention in the last few years, since it is
believed that these models shed light on the solution to
fundamental problems when the universe is traced back to a very
early time. Brane World (WB) cosmology offered a novel approach to
our understanding of the evolution of the universe at higher
dimension. The most spectacular consequence of this scenario is
the modification of the Friedmann equation. When a five
dimensional model is considered the matter describe by a scalar
field or any other field, is confined to a four dimensional brane
while gravity can propagate though the bulk. These kinds of models
can be obtained from a higher superstring theory \cite{horawa}.
For a review on BW cosmology, see Refs.\cite{lindsey} among
others.

Coming from higher dimensional gravity theory, the tachyonic field
matter might provide an explanation for inflation and could
contribute to some new form of cosmological dark matter at late
times. Although, as it was pointed out in Ref.\cite{Sami_taq}, a
homogeneous tachyonic field evolves toward its ground state
without oscillating about this state. Hence, the conventional
reheating mechanisms in tachyonic models does not work
\cite{Kofman}. As we already mentioned, become the alternative
mechanisms such that have proved to be inefficient the
gravitational particle production on instant preheating
\cite{urena,Sami_taq}. In this sense, we believe that the
introduction of the curvaton field plays an important role in the
reheating period associated o the tachyonic inflationary universe
model. Specifically, we explore the curvaton reheating mechanisms
in tachyonic inflationary unverse models with an exponential
potential (i.e. a NO model) by considering BW cosmology. We follow
a similar procedure to that describe in Refs.
\cite{urena,labanda}. As the energy density decreases, the
tachyonic field makes a transition into a kinetic energy dominated
regime bringing inflation to the end. We considered the evolution
of the curvaton field through three different stages. Firstly, it
is assumed that the curvaton field coexists with the inflaton
field during inflation. In this period the tachyonic energy
density is the dominant component (even through the curvaton field
survives the rapid expansion of the universe). The following
stage, i.e., during the kinetic epoch, is that in which the
curvaton mass term becomes important. In order to prevent a period
of curvaton-driven inflation, the universe must still remain
inflaton-driven during this time. When the effective mass of the
curvaton becomes important, the scalar field starts to oscillate
around at the minimum of its potential. Here the energy density,
associated to this field starts to evolve as a non-relativistic
matter. At the final stage, the curvaton field decays into
radiation and then the standard Big-Bang cosmology is recovered.
In order that the model could work the decay of the curvaton field
should occur before the period of nucleosynthesis. Other
constraints may arise depending on he epoch of the decay, which is
governed by the decay parameter.  Here, there are two scenarios to
be possible, depending on whether the curvaton field decays before
or after it becomes the dominant component of the universe.

In the first stage the dynamics of the tachyon field in BW
cosmology is described in the slow-roll over approach
\cite{Sami_taq}. Nevertheless, after inflation, the term   is
negligible compared to the friction term \cite{Guo}. This epoch is
called "kinetic epoch" or "kination", and we will use the
subscript "k" to label the value of the different quantities at
the beginning of this epoch. Even though, this kinetic epoch does
not occur immediately after inflation, since exist an epoch in
between in which,  the tachyonic potential force is negligible
compared to the friction term since the friction coefficient is
proportional to the potential \cite{Guo}.

\section{\label{sec:level1} Inflationary dynamics}

In this section we give a brief description of an inflationary
model in a braneworld scenario, assuming that the tachyonic field
$\phi$ drive inflation.

For the Randall-Sundrum model\cite{randall}, the standard Friedman
equation is modified, and for large energies is given by
\begin{equation}
H^2=\frac{\kappa_0}{3}\rho_{\phi}\left(1+\frac{\rho_{\phi}}
{2\lambda}\right)\approx\frac{\kappa_0}{6}
\frac{\rho_{\phi}^2}{\lambda}\label{key_2},
\end{equation}
where $\lambda$ is the brane tension,  $\rho_{\phi}$ denote the
energy density for the tachyonic field $\phi$. Here, large
energies indicate that $\rho_{\phi}\gg\lambda$. The Hubble factor
is denote by $H$,  and $\kappa_0=8\pi m_p^{-2}$, $m_p$ is the
Planck mass. The energy density for the tachyonic field
$\rho_\phi$ for an effective fluid with energy-momentum tensor
equal to $T_\nu^\mu=diag(-\rho,p,p,p)$ is given by:
\begin{equation}
\rho_\phi=\frac{V(\phi)}{\sqrt{1-\dot{\phi}^2}}, \label{rhoi}
\end{equation}
 and the tachyonic field satisfied the field equation
\begin{equation}
\frac{\ddot{\phi}}{1-\dot{\phi}^2}+3\,H\,\dot{\phi}=-\frac{V'(\phi)}{V(\phi)},
\label{key_1}
\end{equation}
where the dots denote derivative with respect to time, $V(\phi)$
is the effective scalar potential and the prime indicates a
$\phi-$ derivative. The tachyonic potential $V(\phi)$, is such
that it satisfies $V(\phi)\rightarrow 0$ as
$\phi\rightarrow\infty$. Sen \cite{Sen1} has argued that the
qualitative dynamics of string theory tachyon  can be describe by
the exponential potential,
$$
V(\phi)=V_0e^{-\alpha\sqrt{\kappa_0}\phi},
$$
where $\alpha$ and $V_0$  are free parameters.

In the slow roll-over approaches for BW cosmology, the field
equations are:

\begin{equation}
H=\sqrt{\frac{k_0}{6\lambda}}\;V(\phi), \label{eq1}
\end{equation}
and
\begin{equation}
3H\dot{\phi}=-\frac{V'(\phi)}{V(\phi)}=\alpha\sqrt{k_0}.\label{eq2}
\end{equation}

 In this approximation the scale factor is given by \cite{Sami_taq}
\begin{equation}
\frac{a}{a_i}=\exp\left[\beta
t\left(C-\frac{\alpha^2}{6\beta}\;t\right)\right],
\end{equation}
and the scalar field have the expression
\begin{equation}
\phi=-\frac{1}{\alpha\sqrt{k_0}}\ln\left(C-\frac{\alpha^2}{3\beta}t\right),
\end{equation}
 where
$C=\exp(-\alpha\sqrt{k_0}\phi_i)$,
$\beta=\frac{V_0}{\sqrt{6\lambda k_0}}$  and
$\phi(t=t_i=0)=\phi_i$.

The final time of inflation is given by
\begin{equation}
t_{end}=\frac{3\beta}{\alpha^2}\left(C-\frac{\alpha}{\beta\sqrt{3}}\right),
\end{equation}
and we have
\begin{equation}
V_{end}=\frac{V_{0}\alpha}{\beta\sqrt{3}}=\alpha
\sqrt{2\lambda},\;\;\;\dot{\phi}_{end}^2=\frac{1}{3}.
\end{equation}

The number of e-folding $N$ is express in term of $V_i$ and
$V_{end}$ in the form
\begin{equation} V_{end}^2=\left(
\frac{V_i^2}{2N+1}\right),
\end{equation}
where we have used the definition for the number of e-folding,
given by
$$
N=\ln\frac{a}{a_i}=\int_{t_i=0}^t H(t')dt'.
$$

\section{\label{sec:level1} Kinetic epoch}
The dynamic of the curvaton field $\sigma$ though different stages
are described. This allows us to find the constrains upon the
parameters of the model, and thus, to have a viable curvaton
scenario, together with an steep inflationary scenario. We
considered that the curvaton field obeys the Klein-Gordon
equation, and for simplicity, we assumed that the scalar potential
is given by $U(\sigma)=m^2\sigma^2/2$, where $m$ is the curvaton
mass.

 First of all, it is assumed that the  field $\sigma$ coexists
with the tachyonic field $\phi$ during inflation, but the
tachyonic energy density $\rho_{\phi}$ is the dominant component
in relation the curvaton energy density, i.e. $\rho_\phi\gg
\rho_\sigma$.

In the next scenario, the curvaton field oscillated around the
minimum of the effective potential $U(\sigma)$, its energy density
evolves as a non-relativistic matter. During the kinetic epoch the
universe must remain tachyonic-dominated  until this epoch, this
prevents a period of curvaton-drive inflation. In the last
scenario the curvaton decays into radiation and the standard big
bang cosmology is recovered.

In the inflationary regimen is supposed that the curvaton mass
satisfied the condition   $m\ll\,H_f$ and its dynamics is
described in detail in Ref.\cite{urena}.  When the curvaton energy
is equally distributed into kinetic and potential at the beginning
of inflation, i.e. $\dot{\sigma}_i^2\sim\,m^2\sigma_i^2$,  the
condition $m\ll\,H_f$ makes the curvaton field remain constant
under general initial conditions, and we can write $\sigma_f\simeq
\sigma_i$ and $\dot{\sigma}_f=0$, the subscripts $i$ and $f$ are
used to denote the beginning and end of inflation, respectively.

As it was commented above, the dynamic of the curvaton field
coexists with the surviving tachyon field after inflation, whose
energy density is by far the dominant one at the end of inflation.
The hypothesis is that during the kinetic epoch  the Hubble
parameter has decreased so that its value becomes comparable whit
the curvaton field mass $m$. At this time we have $m\simeq H$.

The  dynamics of the BW cosmology for the tachyon field
 in the kinetic regimen is described by the equations:
 \be
\frac{\ddot{\phi}}{1-\dot{\phi}^2}+3\,H\,\dot{\phi}=0,
\label{key_1}
 \en
and \be
 H^2\approx\frac{k_0}{6\lambda}\rho_{\phi}^2.
\en

From Eq.~(\ref{key_1}) we find a first integral for $\dot{\phi}$
in terms of the scale factor given by
\begin{equation}
\dot{\phi}^2=\frac{1}{1+C\,a^6}\,;\,\;\;\;\;\;\;\;
C=\frac{1-\dot{\phi_k}^2}{\dot{\phi_k}^{2}a_k^{6}}\,
>0,\label{firin}
\end{equation}
$C$ is an integration constant, $\dot{\phi_k}$ and $a_k$ represent
values at the beginning of the kinetic epoch for the time
derivative of the tachyonic field and the scale factor,
respectively. A universe dominated by tachyonic field would go
under accelerated expansion if $\dot{\phi}^2<\frac{1}{3}$. The end
of inflation is characterized by the value
$\dot{\phi}_{end}^2=\frac{1}{3}$. By now, the value of
$\dot{\phi}$ at the beginning of the kinetic epoch lies in the
range $1\gtrsim\dot{\phi_k}^2\gtrsim\frac{1}{3}$.

Substituting the expression for $\rho_\phi$, the potential
$V(\phi)$ and $\dot{\phi}^2$ from the above expressions,
Eq.(\ref{firin}), and by using that
$$\dot{a}=\frac{d\,a}{d\,\phi}\,\,\dot{\phi}=
\frac{1}{(1+Ca^6)^{1/2}}\frac{d\,a}{d\,\phi},
$$
 the resulting equation is
integrated for obtaining a relation between the potential $V$ and
the scale factor $a$, given by

$$
 V(\phi)=V(\phi(a))=V=V_0e^{-\alpha\sqrt{k_0}\phi_k}-\alpha\sqrt{\frac{2\lambda}{3}}\times
$$
\begin{equation}
 [ArcTan(\sqrt{C}a^3)-ArcTan(\sqrt{C}a_k^3)],
\end{equation}
and the Hubble factor becomes
\begin{equation}
H=\sqrt{\frac{k_0}{6C\lambda}}\;\;\frac{V}{a^3}\;\sqrt{1+Ca^6}=H_k\frac{V}{V_k}\frac{a_k^3}{a^3}
\frac{\sqrt{1+Ca^6}}{\sqrt{1+Ca_k^6}}.\label{H}
\end{equation}
The energy density in terms of the scale factor results to be
\begin{equation}
\rho_{\phi}=\rho_\phi^k\frac{V}{V_k}\frac{a_k^3}{a^3}\frac{\sqrt{1+Ca^6}}
{\sqrt{1+Ca_k^6}}\label{rho_phi},
\end{equation}
where
\begin{equation}
H_k=\sqrt{\frac{k_0}{6\lambda}}\,\;\rho_{\phi}^k\label{del} .
\end{equation}
When the curvaton mass satisfies $m\simeq\,H$, then we find that
\begin{equation}
\frac{m}{H_k}=\frac{V_m}{V_k}\,\frac{a_k^3}{a_m^3}\,\frac{\sqrt{1+Ca_m^6}}
{\sqrt{1+Ca_k^6}}.\label{H_k}
\end{equation}

In order to prevent a period of curvaton-driven inflation, the
universe must still be dominated by the tachyonic field, i.e.
$\rho_{\phi}|_{a_m}=\rho_{\phi}^{(m)}
\gg\rho_{\sigma}\sim\,U(\sigma_f)\simeq\,U(\sigma_i)$. This
inequality allows us to find a constraint on the initial value of
the curvaton field in the inflationary scenario. Hence, from Eq.
(\ref{key_2}), at the moment when $H\simeq m$, we establish the
restriction

\begin{equation}
\frac{m^2\sigma_i^2}{2\rho_\phi^{(m)}}=
\frac{m\sigma_i^2}{2}\frac{\sqrt{k_0}}{\sqrt{6\lambda}}\ll1
\Rightarrow \sigma_i^2\ll\frac{\sqrt{24\lambda}}{m\sqrt{k_0}}
.\label{pot}
\end{equation}
This expression does not remain as in the case of
Ref.\cite{urena}, due to the presence of the brane tension. This
its becomes subdominant at the time when the curvaton mass $m$ is
the order of $H$. Moreover, the curvaton energy should also be
subdominant at the end of inflation. This restriction justifies
the initial hypothesis in which the curvaton mass results to be
smaller than the Hubble parameter at the end of inflation. This
gives a new result on the constraint of $m$ at different that
happened in the standard cosmology. Now, the ratio between the
potential energies at the end of inflation $U$ and $V$, is given
by

\begin{equation}
\frac{U_f}{V_f}=\frac{m^2\sigma_i^2\sqrt{k_0}}{H_f\sqrt{24\lambda}}\ll\frac{m}{H_f}.
\end{equation}
Here we have used  Eq.(\ref{pot}). In this way, the curvaton mass
should obey the constraint in the tachyonic models

\begin{equation}
m\ll H_f, \label{one}
\end{equation}
condition that sure that curvaton field remains constant under
general initial conditions, i.e. the curvaton energy is equally
distributed, so that $\sigma_f\simeq \sigma_i$. In our case the
value of the curvaton mass does not depend on the parameter
$\alpha$ that appears in the effective potential $V(\phi)$ as in
the case of standard inflation. In the standard case we observe
that the constraint on the curvaton mass results to be increase by
a factor $1/\alpha$. However, in our case, this situation does not
occur, allowing to decreases the strong constraint in the range of
curvaton mass.

After the curvaton field becomes effectively massive, its energy
decays as a non-relativistic matter in the form

\be \rho_\sigma =\frac{m^2\sigma_i^2}{2}\frac{a_m^3}{a^3}
\label{c_cae}. \en

As we have anticipated above the curvaton field cross by different
stages. The curvaton decay provides two different scenarios: one
when curvaton decay after domination and other one when the decay
occurs before dominating.

On the other hand, follows the detailed analysis doing in
\cite{urena} we show that the  produced scalar perturbation
amplitude is related with the other parameters of the model
through the curvaton fluctuations. During the time that the
fluctuations are inside the horizon, they obey the same
differential equation as the inflaton fluctuations do. From that
we may conclude that they acquire an amplitude given by
$\delta\sigma_i\simeq H_i/2\pi$. Once the fluctuations are out of
the horizon, they obey the same differential equation as the
unperturbed curvaton field. then we expect that they remain
constant during inflation, under quite general initial conditions.
Now we should take into account that the tachyon field survives
the inflationary period, and study how the final spectrum of
perturbation could be modify by this.

For the first scenario, when the curvaton comes to dominate the
cosmic expansion, there must be a moment when the tachyonic  and
curvaton energy densities becomes equal. Joining the expressions
for the ratio of energy densities and the value of the Hubble
parameter, we find the Hubble parameter as a function of the
parameters introduced by the curvaton. This gives us a new result
for the range of the decay parameter. This parameter has an upper
limit which becomes smaller than its analogous value comparing to
the standard case. that means a diminishing in the reheating
temperature as one will see forward. From Eqs. (\ref{rho_phi}),
(\ref{del}), (\ref{H_k}) and (\ref{c_cae}) at the times when
$(\rho_\sigma=\rho_\phi)$, we get
$$
\left.\frac{\rho_\sigma}{\rho_\phi}\right|_{a=a_{eq}}=\frac{\sqrt{k_0}m^2\sigma_i^2}{\sqrt{24\lambda}H_k}
\frac{a_m^3}{a_k^3}\frac{V_k}{V_{eq}}\frac{\sqrt{1+Ca_k^6}}{\sqrt{1+Ca_{eq}^6}}=
$$
\begin{equation}
  =\frac{\sqrt{k_0}}{\sqrt{24\lambda}}
\frac{V_m}{V_{eq}}\frac{\sqrt{1+Ca_m^6}}{\sqrt{1+Ca_{eq}^6}}\;m\sigma_i^2
=1.\label{equili}
\end{equation}
Now from Eqs.(\ref{H}),(\ref{H_k})and(\ref{equili}), we find  a
relation between the Hubble parameter, at the moment when the
energy densities are the same $H_{eq}$, in terms of curvaton
parameter and the ratio of the scale factor at different epochs,
given by:

\begin{equation}
H_{eq}=H_k\frac{V_{eq}}{V_k}\frac{a_k^3}{a_{eq}^3}\frac{\sqrt{1+Ca_{eq}^6}}{\sqrt{1+Ca_{k}^6}}
=\sqrt{\frac{k_0}{24\lambda}}\,\;\frac{a_m^3}{a_{eq}^3}\,m^2\sigma_i^2.
\label{heq}
\end{equation}

The decay parameter $\Gamma_\sigma$ is constrained by the
requirement that the curvaton field decay before of
nucleosynthesis happen. Then, we have that $H_{nucl}=10^{-40}m_p <
\Gamma_\sigma$. On the other hand, we also require that the
curvaton decays after domination, and thus $\Gamma_\sigma <
H_{eq}$ so that we obtain a constrain on the decay parameter,
given by

\begin{equation}
10^{-40}m_{p}<\Gamma_{\sigma}<\sqrt{\frac{k_0}{24\lambda}}\,\;
\frac{a_m^3}{a_{eq}^3}\,m^2\sigma_i^2 \label{gamm1}
\end{equation}
with $a_m <a_{eq}$. In the standard case
$H_{eq}^{(std)}=(3m_p^2)^{-1}4\pi \sigma_i^2 m$ thus, \be
\Gamma_\sigma\,<\,H_{eq}<\;\frac{m\,m_p}{\sqrt{\lambda}}\,H_{eq}^{(std)}.
\en Note that for the tachyon model in the BW cosmology the range
of $\Gamma_\sigma$ is shorter than the standard case.

Now it is interesting to give an estimation of the constrain on
the parameters of the our model, using the scalar perturbation
related to the curvaton field. We would like to point out here
that the primordial curvature perturbation may have a completely
different origin, namely the quantum fluctuation during inflation
of a light scalar field (the curvaton scalar field)which is not
the slowly-rolling inflaton, and need have nothing to do with the
fields driving of inflation. In general, we may say that the
curvaton creates the curvature perturbation in two separate
stages. In the first stage, its quantum fluctuation during
inflation is converted at horizon exit to a classical perturbation
with a flat spectrum. Then in its second stage, after inflation,
the perturbation in the curvaton field is converted into a
curvature perturbation. At different with the usual mechanism, the
generation of curvature by the curvaton requires no assumption
about the nature of inflation, beyond the requirement that the
Hubble parameter is practically constant. Instead, it requires
certain properties of the curvaton and of the cosmology after
inflation so that the required curvature perturbation will be
generated. In this way, the evolution of the curvaton fluctuations
resembles that of previous scenarios already present in the
literature \cite{ref1u,urena}. The spectrum of the Bardeen
parameter $P_\zeta$, whose observed value is about $2\times
10^{-9}$, allows to determine an initial value of curvaton field
in terms of the parameter $\alpha$. At the time when the decay of
the curvaton fields occur, the Bardeen parameter becomes
\cite{urena}

\begin{equation} P_\zeta\simeq \frac{1}{9\pi^2}\frac{H_i^2}{\sigma_i^2}.
\label{pafter}
\end{equation}
The spectrum of fluctuations is automatically gaussian since
$\sigma_i^2>>H_i^2/4\pi^2$, and is independent of $\Gamma_\sigma$,
a feature that will simplify the analysis on the parameter space.
Moreover, even though the curvaton coexists with the inflaton
field up to this epoch, the spectrum of fluctuations is the same
as in the standard scenario.

From Eq.(\ref{pafter}) and by using that
$H_i^2=(k_0/6\lambda)\,(2N+1)V_f^2$ and $V_f^2=2\alpha^2\,\lambda$
(reported  in Ref.\cite{Sami_taq}) we relate the perturbation with
the parameters of the model in such way that

\begin{equation}
\frac{27\pi}{4}\frac{P_{\zeta}}{(2N+1)}\;\sigma_i^2=\frac{\alpha^2}{m_p^2}.
 \label{18}
\end{equation}
This expression allows us to fix the initial value of the curvaton
field in term of the  free parameter $\alpha$. By using
Eq.(\ref{18}), the constraint Eq.(\ref{one}) transformed into the
following constraint for the curvaton mass

\begin{equation}
\frac{m}{m_p}\ll \frac{3\pi
P_\zeta^{1/2}}{(2N+1)^{1/2}}\frac{\sigma_i}{m_p}.
\end{equation}

Finally, Eq.(\ref{gamm1}) restricts the value of the decay
parameter $\Gamma_\sigma$, which can be transformed into another
constraint upon $m$, $\sigma_i$ and $\lambda$, such that

\be \frac{m^2\;\sigma_i^2}{\sqrt{\lambda}}\gg 10^{-40} m_p^2.\en

On the other hand, considering that the curvaton field decays
before it dominates the  cosmological expansion (which we called
second scenario), but after becomes   $m\sim H$. The curvaton
decays at a time when  $\Gamma_\sigma =H$ and then from
Eq.(\ref{H}) we get

\begin{equation}
\frac{\Gamma_\sigma}{H_k}=\frac{V_d}{V_k}\frac{a_k^3}{a_d^3}
\frac{\sqrt{1+Ca_d^6}}{\sqrt{1+Ca_k^6}},
\label{Gamm}
\end{equation}
where ` d' labels the different quantities at the
time of curvaton decay.

The curvaton field should decay after it becomes $m\sim H$, so
that $\Gamma_\sigma<m$; and before it dominates the expansion of
the Universe, $\Gamma_\sigma>H_{eq}$ (see Eq.(\ref{heq})).
Therefore, we have

\begin{equation}
\sqrt{\frac{k_0}{24\lambda}}\,\;\frac{a_m^3}{a_{eq}^3}
\,m^2\sigma_i^2\,<\Gamma_\sigma<m.
\label{gamm2}
\end{equation}

As we discussed above the curvaton field decays before it
dominates the cosmological expansion, but after it becomes massive
the slower limit is more closer to decay parameter than in the
standard case as we could see from the follow expression

\be
m>\Gamma_\sigma^{(std)}>H_{eq}^{(std)}>\sqrt{\frac{\lambda}{m^2m_p^2}}\;H_{eq},
\en
since we have that $a_m<a_{eq}$.  In this form, we could predict
that the curvaton field should decay long before it dominates the
expansion of the universe in comparison with the standard model,
since $H_{eq}^{(std)}>H_{eq}^2/m$ .

 Now for the second scenario, the curvaton decays at the time
when $\rho_\sigma<\rho_\phi$. If we defined the $r_d$ parameter as
the ratio between the curvaton and the tachyonic energy density,
evaluated at $a=a_d$ and with the help of Eq.(\ref{Gamm})

$$
r_d=\left.\frac{\rho_\sigma}{\rho_\phi}\right|_{a=a_d}=\frac{m^2\sigma_i^2\sqrt{k_0}}
{H_k\sqrt{24\lambda}} \frac{V_{k}a_m^3\sqrt{Ca_{k}^6+1}}
{V_{d}a_{k}^3\sqrt{Ca_{d}^6+1}}=
$$
\begin{equation}
=\frac{m^2\sigma_i^2\sqrt{k_0}}{\Gamma_\sigma\sqrt{24\lambda}}
\frac{a_m^3}{a_d^3} \label{rd},
\end{equation}
 and for $r_d\ll 1$ (in agreement with the Eq.(\ref{gamm2})),
the Bardeen parameter is given by \cite{ref1u}

\begin{equation} P_\zeta\simeq \frac{r_d^2}{36\pi^2}\frac{H_i^2}{\sigma_i^2}.
\label{pbefore}
\end{equation}

When the curvaton decay before domination the expression
(\ref{pbefore}) could be write as \be
\frac{\sigma_i^2}{m_p^2}=108\pi\;\frac{\lambda}{m^4}
\left(\frac{a_d}{a_m}\right)^6\frac{P_\zeta}{(2N+1)}
\left(\frac{\Gamma_\sigma}{H_f}\right)^2 ,\en

now the expression (\ref{gamm2}) is write as

\be
108\sqrt{\frac{\lambda\pi^3m_p^2}{3m^4}}\frac{a_d^6}{a_m^{3}a_{eq}^{3}}
\frac{P_\zeta}
{\;(2N+1)}\left(\frac{\Gamma_\sigma}{H_f}\right)^2<\Gamma_\sigma<m,
\en and from the above inequality, we find

\begin{equation}
\Gamma_\sigma < \sqrt{\frac{3m^4}{\lambda \pi^3
m_p^2}}\frac{(2N+1)H_f^2}{108 P_\zeta}.
\end{equation}
Using the values $P_\zeta\sim 10^{-5}$, $\lambda\sim
10^{-10}m_p^4$ and $H_f\sim 10\alpha /m_p$, we find that
\begin{equation}
\Gamma_\sigma< \left(\frac{10^{6}\alpha m}{
m_p^{5/2}}\right)^2,\label{rr}
\end{equation}
which can be used to estimate an upper limits on $\Gamma_\sigma$
in term of $\alpha$, as well as $m$.

In the first scenario, our calculations allow us to get the
reheating temperature as hight as $10^{-5}m_p$, since the decay
parameter $\Gamma_{\sigma}\propto\,T_{rh}^2/m_p$, where $T_{rh}$
represents the reheating  temperature. Here, we have used Eqs,
(\ref{gamm1}) and (\ref{18}), $a_m/a_{eq}\sim\,10^{-1}$, $m\sim
10^{-5}m_p$, $\lambda\sim 10^{-10}m_p^4$ and the value
$\alpha\sim\,10^{-4} m_p^2$ \cite{labanda,Dimo}.

In the second scenario we have that, from the Eq. (\ref{rr}), we
could estimate the reheating temperature to be of the order of
$\sim 10^{-3}m_p$ as an upper limit.

As it was reported in Ref.\cite{Sami_taq}, at the end of inflation
$\rho_\phi$ at best could scales as $a^{-3}$, which is valid
irrespectively of the form of the tachyonic potential, providing
that it satisfies $V(\phi)\rightarrow 0$ as $\phi\rightarrow
\infty$. However, this is not true in general, since in our
particular case, we have found that it is possible to get a more
complex expression for the dependence of the energy density
$\rho_\phi$ in terms of the scale factor, how we could see from
Eq. (\ref{rho_phi}).

We should mention that, we have introduced the curvaton mechanism
into NO inflationary brane tachyonic model  as another possible
solution to the problem of reheating, where there is not need to
introduce an interaction between the tachyonic and some auxiliary
scalar field.

\begin{acknowledgments}
CC was supported by MINISTERIO DE EDUCACION through MECESUP Grants
FSM 0204. SdC was also supported by COMISION NACIONAL DE CIENCIAS
Y TECNOLOGIA through FONDECYT Grants N$^0$ 1030469, 1040624 and
1051086, and by UCV-DGIP N$^0$ 123.764.

\end{acknowledgments}


\end{document}